\documentclass[
reprint,
superscriptaddress,
amsmath,
amssymb,
aps,
prb,
]{revtex4-2}
\usepackage{ragged2e}
\usepackage{caption}
\usepackage[caption=false]{subfig}
\usepackage{ragged2e}
\DeclareCaptionJustification{justified}{\justifying}
\usepackage[colorlinks,allcolors=cyan]{hyperref}
 
\usepackage{graphicx}
\usepackage{dcolumn}
\usepackage{bm}
\usepackage[capitalise]{cleveref}
\usepackage{physics}
\usepackage{xcolor}
\usepackage{comment}
\usepackage{bbold}
\usepackage{mathtools}
\usepackage{dsfont}
\usepackage{bbold}
\usepackage[caption=false]{subfig}
\DeclareCaptionOption{justified}[]{\caption@setjustification{justified}}

\begin{document}
\def\lc{\left\lfloor}   
\def\rc{\right\rfloor}
\setlength{\intextsep}{10pt plus 2pt minus 2pt}

\title{Stochastic Sampling of Operator Growth Dynamics}
 
\author{Ayush De}
 \affiliation{Racah Institute of Physics, The Hebrew University of Jerusalem, Jerusalem 9190401, Israel}
 \author{Umberto Borla}
 \affiliation{Racah Institute of Physics, The Hebrew University of Jerusalem, Jerusalem 9190401, Israel}
\author{Xiangyu Cao}
\affiliation{Laboratoire de Physique de l'\'Ecole normale sup\'erieure, ENS, Universit\'e PSL, CNRS, Sorbonne Universit\'e, Universit\'e Paris Cit\'e, F-75005 Paris, France}
 \author{Snir Gazit}
 \affiliation{Racah Institute of Physics, The Hebrew University of Jerusalem, Jerusalem 9190401, Israel}
 \affiliation{The Fritz Haber Research Center for Molecular Dynamics, The Hebrew University of Jerusalem, Jerusalem 9190401, Israel}
\date{\today}

\begin{abstract}
We put forward a Monte Carlo algorithm that samples the Euclidean time operator growth dynamics at infinite temperature. Crucially, our approach is free from the numerical sign problem for a broad family of quantum many-body spin systems, allowing for numerically exact and unbiased calculations. We apply this methodological headway to study the high-frequency dynamics of the mixed-field quantum Ising model (QIM) in one and two dimensions. The resulting quantum dynamics display rapid thermalization, supporting the recently proposed operator growth hypothesis. Physically, our findings correspond to an exponential fall-off of generic response functions of local correlators at large frequencies. Remarkably, our calculations are sufficiently sensitive to detect subtle logarithmic corrections of the hypothesis in one dimension. In addition, in two dimensions, we uncover a non-trivial dynamical crossover between two large frequency decay rates. Lastly, we reveal spatio-temporal scaling laws associated with operator growth, which are found to be strongly affected by boundary contributions.
\end{abstract}
\maketitle
\section{Introduction}
Within the Heisenberg representation of quantum mechanics, the time evolution of operators is governed by their commutation relations with the Hamiltonian. Under generic many-body dynamics, local operators are expected to evolve into increasingly complex and non-local operator strings. This process of ``operator growth'' is equivalent to the scrambling of quantum information in complex systems and is closely related to quantum chaos. As such, understanding the precise properties of operator growth is a fundamental line of research that has driven a great amount of theoretical and experimental work~\cite{measuring-scramcling,nahum-op-spreading,khemani-RUC-conserve,khemani-huse-nahum,gopalakrishnan,Qistreicher,growthsyk,scrambling-xu,swingle-review}. It has direct links to the question of how the laws of quantum mechanics bring about thermalization or its failure in isolated systems~\cite{deutsch,srednicki,rigol-review,colloque-thermo,huse-nandik-review}. It is also a useful tool for investigating the fate of information in a black hole and how to retrieve it~\cite{Sekino_2008,shenkerstanford,boundonchaos,brown19-teleport,schuster22-tele}. 

In this context, an interesting way to quantify the rate of operator growth, besides the much studied out-of-time order correlator, is to examine the high-frequency regime of the spectral density $\Phi(\omega)$ associated with generic dynamical correlators~\cite{elsayed-fine}. Recently, Ref. \cite{Parker_2019} proposed an operator growth hypothesis that translates to an exponentially decaying large frequency tail
\begin{equation}\label{eq:phi}
    \Phi(\omega) \sim \exp(- |\omega| / \omega_0) 
\end{equation}
in generic many-body systems with finite local bandwidth. The energy scale $\omega_0$ provides an upper bound to the Lyapunov exponent of out-of-time order correlators~\cite{Parker_2019,Murthy-Srednicki,zhang-gu-kitaev}, and is thus an alternative measure of operator growth~\cite{rabin-0,rabin-1,rabin-2,Dymarsky_2020,dymarsky,Avdoshkin2022,swingle-kcomp}. Crucially, this quantity is experimentally accessible~\cite{floquet-bloch}, e.g., from the pre-thermalization time in a fast periodic drive~\cite{ADHH}.

\begin{figure}[h]
\captionsetup[subfigure]{labelformat=empty}
    \subfloat[\label{subfig:conf_ex}]{}
    \subfloat[\label{subfig:op_den}]{}
     \includegraphics[width=\columnwidth]{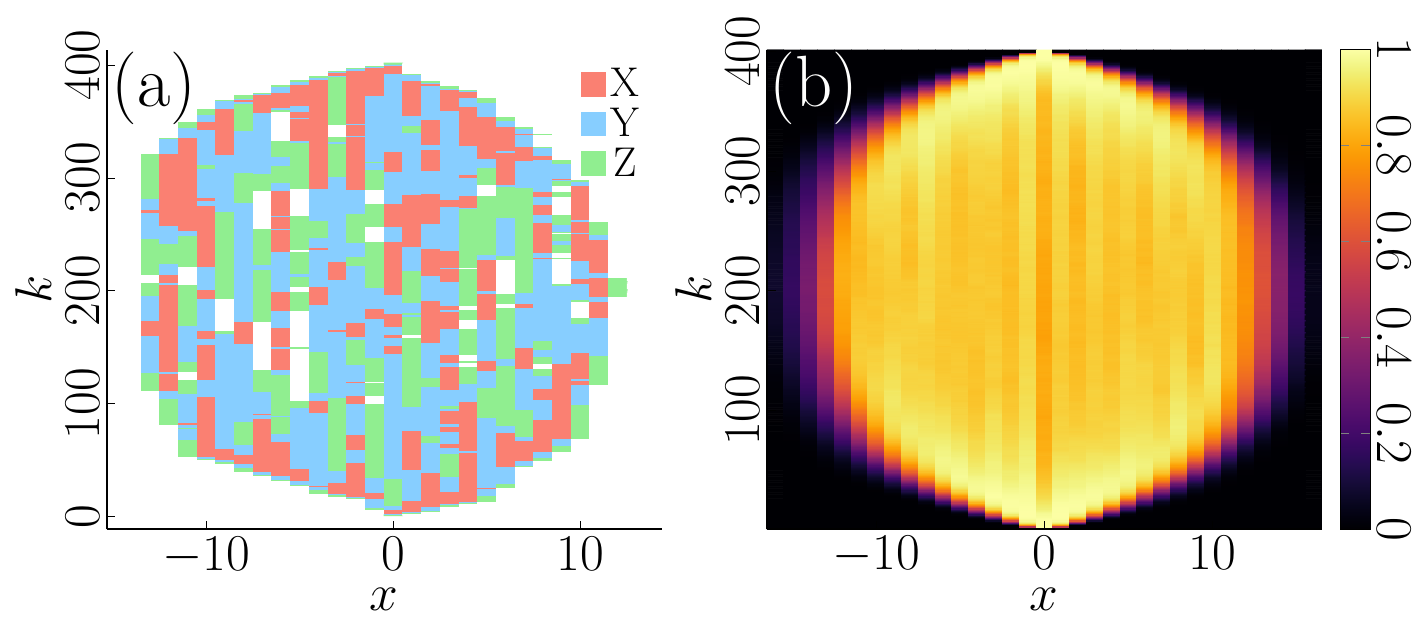}

\caption{\label{fig:og} (a) The operator growth history of the $Z$ Pauli operator associated with a typical sequence belonging to $\mathcal{S}_{2n}$ for $n=200$ in \cref{eq:musum} in the Pauli basis, for the mixed field QIM with Hamiltonian given in \cref{eq:hamiltonian}. The horizontal axis corresponds to space, and the vertical axis is generated by the temporal index $k$. The colors correspond to different Pauli operators at each site. (b) The average Pauli operator density across growth histories contributing to $\mu_{2n}$ for $n=200$. Here, we only distinguish between the non-trivial Pauli operators and the identity. Both (a) and (b) have been extracted from our Monte Carlo simulations.}
\end{figure}

Studying operator growth dynamics in concrete lattice models is particularly challenging due to the inherent difficulty of simulating many-body quantum dynamics with classical resources, as reflected by the dynamical sign problem and the unbounded growth of entanglement entropy in thermalizing systems \cite{Kim_2023, Muhlbacher_2008}. From the perspective of the operator growth hypothesis, direct evidence is largely limited to large-N or semi-classical models~\cite{zhang-gu-kitaev,swingle-kcomp,largeq-nandy,elsayed-fine,semiclassic} and brute-force numerics of small to intermediate system sizes that suffer from sizable finite size effects~\cite{Gemmer_2022}. Although analytical bounds supporting the conjecture have been obtained for finite-dimensional systems~\cite{Bouch,Cao_2021}, these results are not satisfactory in that they do not provide an accurate estimate of $\omega_0$ or the precise functional form of the spectral function. It is, therefore, desirable to devise efficient numerical schemes that can reliably test the operator growth hypothesis. 

In this article, we present a Monte Carlo method that samples the Euclidean time operator growth dynamics at infinite temperature in an unbiased and statistically exact manner. A unique feature of our formalism is that it tracks the Euclidean time evolution of \textit{operators} (\cref{subfig:conf_ex}) instead of the standard wave function approach that is common in conventional Quantum Monte Carlo (QMC) calculations. Crucially, our scheme does not suffer from the notorious numerical sign problem when applied to a broad family of lattice spin models of interest. By way of example, we employ our method to study the mixed-field Quantum Ising Model (QIM) in one- and two-dimensions and succeed in computing an order of magnitude more number of spectral function moments than what is accessible using brute force methods. This progress gives us direct access to high-frequency decay rates of the spectral tail. Furthermore, our results provide high-precision tests of the operator growth hypothesis. In particular, we pinpoint the subtle log-correction to \cref{eq:phi} in 1D and unveil non-universal spectral features like an intriguing crossover between two high-frequency tails in 2D. Additionally, we determine the dynamical scaling properties of the operator support size.

\section{Constructing the Operator-Space Partition Function}
In statistical physics, Monte Carlo schemes aim to tackle exponentially hard computational problems by mapping them to a sampling problem from a probability distribution. A useful starting point for constructing our partition function is to examine the auto-correlation function of a local operator, $O$, analytically continued to imaginary time $t=i\tau$: 
\begin{equation} \label{eq:Ctau}
    \mathcal{C}(\tau)=\langle O(\tau)O(0)\rangle=( O \vert e^{\mathcal{L} \tau}  |O ).
\end{equation} 

Here, $( A \vert B ) = \mathrm{Tr}[A^\dagger B] / \mathrm{Tr}[1]$ denotes the infinite-temperature scalar product on operator space. This enables a natural extension of the \textit{bra-ket} notation to the operator Hilbert space, with $|O)$ denoting the operator $O$ and $(O|$ its vector dual $O^{\dagger}$. $\mathcal{L}$ here is the Liouvillian super-operator associated with Hamiltonian $H$ that transforms operator \textit{kets} as $\mathcal{L}|O)=|[H, O])$.

 To make contact with the real frequency spectral density (and especially its large frequency tails) we Taylor expand $\mathcal{C}(\tau)$ about $\tau = 0$:
\begin{align}
     &\mathcal{C}(\tau)  = \sum_{n=0}^{\infty} \mu_{2n} \frac{\tau^{2n}}{(2n)!}  \label{eq:Ctau-exp} \\  
     & \mu_{2n} = ( O  | \mathcal{L}^{2n} | O ) = \int  \Phi(\omega) \omega^{2n} \frac{\mathrm{d} \omega}{2\pi} \label{eq:mu2n}.
\end{align}
In the last equation, we identify the Taylor expansion coefficients  $\{\mu_{2n}\}$ with the moments of the spectral function $\Phi(\omega)$, defined as $\Phi(\omega)=\int d\tau e^{i \omega \tau} \langle O(\tau=it)O(0)\rangle $. The moments thus encode information about the dynamics of the operator. An example of the high frequency part of spectral density associated with dynamics with the $Z$ Pauli operator for the mixed-field QIM Hamiltonian \cref{eq:hamiltonian} can be found in  \cref{eq:twotails}. In particular, an exponential tail of the spectral function (\cref{eq:phi}) yields:

\begin{equation}
    \mu_{2n} \simeq \left( \frac{2 n \omega_0}{e} \right)^{2n}
    \label{eq:large_mu}
\end{equation}
for large $n$. The moments grow rapidly, such that $\mathcal{C}(\tau)$ converges only when $\tau < \tau_* = 1/\omega_0$, and has a pole at $\tau_*$. 

We aim to extract dynamical information by evaluating the moments. To map this calculation to a statistical mechanics problem, we express $\mu_{2n}$ as a sum over paths in operator space. Assuming local dynamics, we write $\mathcal{L}=\sum_{i,a}\mathcal{L}_{i,a}$ , with  $\left\{\mathcal{L}_{i,a}\right\}$ denoting local Liouvillian of type $a$ acting on site $i$, as dictated by the underlying Hamiltonian, $H=\sum_{i,a} H_{i,a}$, with $\mathcal{L}_{i,a}|O)=|[H_{i,a},O])$. We then expand $ ( O  | \mathcal{L}^{2n} | O )  $ by introducing operator space resolution of identities between individual Liouvillians:
\begin{equation} \label{eq:musum}
    \mu_{2n} = \sum_{\mathcal{S}_{2n}} \underbrace{\prod_{k=0}^{2n-1} (O_{k+1}|\mathcal{L}_{i_k,a_k} O_{k}) }_{ W_{\mathcal{S}_{2n}}} \,,
\end{equation}
where the sum is over all local Liouvillian sequences, specified by $2n$ position-type pairs, $\mathcal{S}_{2n}= \{(i_k,a_k)\}_{k=0}^{2n-1}$, with nonzero weight $W_{\mathcal{S}_{2n}} \ne 0$. This construction implicitly defines operator histories between orthogonal operators $(O_0 = O, O_1, \dots, O_{2n} = O_{0})$. \cref{eq:musum} imparts moments $\mu_{2n}$ the interpretation of a sum over all $2n$ long operator string \textit{paths} generated by local Liouvillian sequences $S_{2n}$ such that $O_{0}=O_{2n}=O$. \cref{fig:og} illustrates this interpretation. We now consider the sum over all such operator paths up to $2n_{\max}$ in length:

\begin{equation}
\label{eq:G}
G=\sum_{n=0}^{n_{\max}}\mu_{2n}=\sum_{n=0}^{n_{\max}}\sum_{\mathcal{S}_{2n}} \prod_{k=0}^{2n-1} (O_{k+1}|\mathcal{L}_{i_k,a_k} O_{k}).
\end{equation}

One can view $G$ as a partition function associated with a probability distribution on $n$ with $P(n)=\mu_{2n}/G$. In that case, the relative frequencies of sampling different values of $n$ from $P(n)$ correspond to the relative magnitudes of the moments. This forms the central idea behind our sampling approach to computing the moments.

To tailor $G$ for numerical Monte Carlo calculations, we make additional adjustments. Firstly, we make each operator string path $2n_{\max}$ long by inserting additional factors of identity Liouvillians. This results in duplication of paths which is compensated for by dividing by a binomial combinatorial factor. This step is reminiscent of the Stochastic Series Expansion (SSE) \cite{Sandvik_1990, Sandvik_2010}, except that here, the configurations are in operator space with transitions being generated by local Liouvillians as opposed to local Hamiltonians. Also in contrast to SSE, where the probability distribution of the expansion order rapidly decays away from its mean value (associated with the energy), here we will be interested in sampling all moments $\mu_{2n}$ and thus drawing samples from all orders of the expansion. \cref{eq:large_mu} however suggests that the moments grow rapidly with $n$. Naively sampling from $G$ would thus result in a highly skewed sampled histogram, with almost all samples coming from the large expansion orders close to $n_{\max}$. This would result in poor statistics for small to intermediate expansion orders. To counter this, we reweigh the moments by numerically determined factors $w_{n}$ such that $w_{n}\mu_{2n}$ is of the same order of magnitude for all $n$. Accounting for the reweighting and the insertion of identity Liouvillians, our final generalized partition function now takes the form:
\begin{equation}
\label{eq:Z}
\mathcal{Z}=\sum_{n=0}^{n_{\max}}w_{n} \underbrace{\sum_{\mathcal{S}_{2n}'} \frac{1}{{2n_{\max} \choose 2n}}\prod_{k=0}^{2n_{\max}-1} (O_{k+1}|\mathcal{L}_{i_k,a_k} O_{k})}_{\mu_{2n}}.
\end{equation}
Here, now, $\mathcal{S}_{2n}'$ sums over all possible $2n_{\max}$ long Liouvillian sequences with $2n-2n_{\max}$ Liouvillian identities and ${2n_{\max} \choose 2n}$ is the aforementioned combinatorial binomial factor, compensating for the over counting introduced by identity Liouvillians. 

It should be noted that in $n_{\max}$ steps of commutation with a $k$ local Hamiltonian, an operator can spatially grow by at most $(k-1) \times n_{\max}$ sites. Thus, for an operator $O$ with finite support, a lattice with a linear spatial size larger than the sum of the support size of $O$ and $(k-1) \times n_{\max}$ avoids any finite size effects in the calculation of the first $n_{\max}$ moments.

An interesting observation is that for a particular choice of weights $w_{n}=\tau^{2n}/(2n)!$, $\mathcal{Z}$ becomes the truncated Taylor expansion of $\mathcal{C}(\tau)$ from \cref{eq:Ctau-exp}. For large $n_{\max}$, a choice of weights $w_{n}$ that ensures a flat distribution across $n$ would correspond to $\mathcal{C}(\tau)$ near criticality, $\tau \nearrow \tau_*$.

\begin{figure}
\includegraphics[width=\columnwidth]{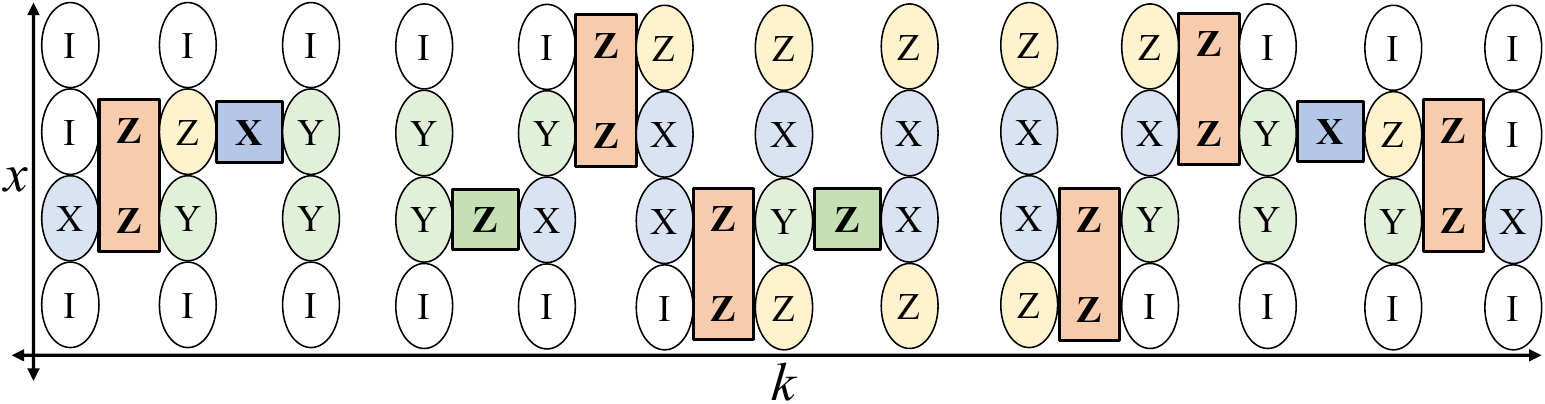}
\vspace{.1cm}
\caption{\label{fig:config} A typical configuration contributing to $\mathcal{Z}$ in \cref{eq:Z} with $n=4$, $n_{\max}=5$ and $L=4$ for the 1D mixed field QIM with Hamiltonian in \cref{eq:hamiltonian}. The horizontal and vertical axes correspond to the temporal index $k$ and the spatial index $i_{k}$ respectively. The sites are populated by the $X$, $Y$, and $Z$ Pauli operators or the $I$ (identity) operator. The temporal evolution is generated by Liouvillians of type $\mathcal{L}_{Z}$, $\mathcal{L}_{ZZ}$ and $\mathcal{L}_{X}$ arising from the longitudinal field, bond and transverse field terms in the Hamiltonian respectively. The empty Liouvillian slots correspond to identity Liouvillians.}

\end{figure}

The weights $W_{\mathcal{S}}$ in \cref{eq:musum} arise out of a product of commutations of operators and are thus generally non-positive even when the associated Hamiltonian problem is sign-problem free. Remarkably, for a class of nontrivial models, the sum in \cref{eq:musum} contains strictly non-negative real terms, $W_\mathcal{S}\ge0$ \cite{Cao_2021}, see further details in \cref{sec:sign_prob}. This class includes the mixed-field QIM, with the Hamiltonian:

\begin{equation}
\label{eq:hamiltonian}
H=-J\sum_{\left\langle {i,j}\right\rangle}Z_{i}Z_{j} -\sum_{i}(h^{Z}Z_{i}+h^{X}X_{i}) \,.
\end{equation}
Here, $\left\langle {i,j}\right\rangle$ define nearest neighbours on a hypercubic lattice and $X_i, Z_i$ are the Pauli operators acting on sites $i$ along the $x, z$ spin axes, respectively. For such spin $1/2$ models, Pauli strings, i.e. operators formed by the tensor product of local Pauli operators, are a convenient orthogonal basis choice for the resolution of Identities in \cref{eq:musum}. To avoid the sign problem, it is sufficient to assume that operator $O_{0}$ is itself a Pauli string. This then imparts the configurations in \cref{eq:Z} a natural diagrammatic interpretation as a sum over configurations on a $D=d+1$ dimensional space-time lattice with dimensions $L^{d}\times (2n_{\max}+1)$. The lattice consists of Pauli strings spanning the spatial dimensions with the temporal dimension generated by local Liouvillians connecting Pauli strings. See \cref{fig:config}.

\section{Sign Problem Resolution}
\label{sec:sign_prob}
\begin{figure}
\centering
\includegraphics[width=100px]{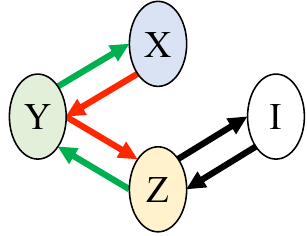}
\vspace{.5cm}
\caption{\label{fig:sign_problem} The \textit{operator state graph} of a site in the mixed field QIM Hamiltonian. The different states correspond to Pauli matrices, transitions between which are generated by local Liouvillians. The red and green arrows depict transitions arising from commutators with $+i$ and $-i$ phases respectively.  The black arrows indicate transitions arising from matrix multiplications as can arise from $\mathcal{L}_{ZZ}$ commutators and thus have no associated phase.}

\end{figure} 

To understand the resolution of the sign problem, it is useful to consider the \textit{operator state graph} for a single site on the Pauli string lattice. \cref{fig:sign_problem} illustrates the same for the mixed field QIM. A site starts off in the state corresponding to the initial Pauli string $O_{0}$. As the Pauli string evolves, the site traces a path in the  \textit{operator state graph}. The boundary conditions imposed in \cref{eq:musum} however demand that each site return to the same state it began in. It can be seen from \cref{fig:sign_problem} that in the mixed field QIM, for a site to return to the same state, it must encounter the same number of red transitions as green transitions, each associated with a phase factor of $+i$ and $-i$, respectively. Consequently, all operator paths will have the same number of $+i$ and $-i$ phase factors,  cancelling out. Thus, all paths will have positive associated weights $W_{\mathcal{S}}$.

More generally, the sign problem is prevented as long as there are no closed loops generated by red or green transitions in the \textit{operator state graph}. In the specific case of the mixed field QIM, shown in \cref{fig:sign_problem}, adding a $h^{Y}Y_{i}$ term to the Hamiltonian will enable additional red and green transitions between states $X$ and $Z$ (via commutations with the $Y$ Pauli matrix). This will allow closed paths of the kind $X \rightarrow Y \rightarrow Z \rightarrow X$ with an $i^{3}$ phase to satisfy the fixed boundary conditions. A sufficient condition to prevent such closed loops (and consequently avoid the sign problem) is thus not to have all three Pauli matrices feature in the Hamiltonian expression. Beyond the mixed field QIM (which encompasses the Transverse Field Ising Model), the quantum XY model (in the absence of a $Z$ field term) is another notable quantum spin model that does not exhibit a sign problem. On the contrary, the quantum Heisenberg model, with its bond terms involving all three Pauli matrices, will suffer from the sign problem.

\section{Numerical methods and observables}
We now present a Monte Carlo sampling scheme for configurations in \cref{eq:Z} for the mixed-field QIM. The moves are designed to traverse the configuration space by making local changes to the existing configuration. We describe them briefly here, and refer to Appendix~\ref{app:MC} for technical details:
\begin{enumerate}
    \item  \texttt{add-drop} --  allows switching between different $n$ sectors by adding or removing a pair of identical local Liouvillians located on the same sites at two consecutive time slices.
    \item  \texttt{swap} --  swaps two commuting Liouvillians at different time slices, without changing the moment order $n$. This move is proposed only if there are no non-commuting Liouvillians at intermediate time slices.
    \item \texttt{interact} --  here, two temporally consecutive $\mathcal{L}_{ZZ}$ Liouvillians that share a single spatial site ``interact" to create a pair of temporally consecutive $\mathcal{L}_{Z}$ Liouvillians acting on next to nearest neighbour sites. This move allows accessing configurations with an odd number of $\mathcal{L}_{ZZ}$ and $\mathcal{L}_{Z}$ Liouvillians acting on a given site.
\end{enumerate}
The weights $w_{n}$ are determined using a reweighting approach similar to the Wang-Landau algorithm \cite{Landau_2001, Troyer_2003, Pereyra_2008}. An initial guess choice is iteratively tuned until reaching an approximately uniform distribution across the different $\mu_{2n}$ sectors. See Appendix~\ref{app:MC} for implementation details. 

Our key physical observable is the moments' ratio:
\begin{equation}
\label{eq:rn_sq}
r_n^2 := \frac{ \mu_{2n+2} }{ \mu_{2n}}
  = \frac{\left\langle \delta_{n',n+1}\right\rangle_{n'}}{\left\langle \delta_{n',n}\right\rangle_{n'}}\times \frac{w_{n}}{w_{n+1}}
\end{equation}

In the above equation,  averaging is carried out over the probability distribution $P(n')$ induced by $\mathcal{Z}$, where the Kronecker delta  $\delta_{n',n}$ counts Monte Carlo events satisfying $n'=n$. The large frequency behavior of \cref{eq:phi}, predicts  a linear growth of $r_{n}$ with $n$, i.e. $r_{n} \simeq 2 \omega_{0}n$.
In addition, the prefactor allows for a numerical estimation of the decay rate $\omega_0$.  

To estimate the spatial extent of operator growth, we define the support size at time slice $k$ as:
\begin{equation}
\label{eq:support}
    S_{k} = \langle \text{Volume}(B^{d}_{k}) \rangle.
\end{equation}
Here, $B^{d}_{k}$ is geometrically defined as the largest region in $d$ dimensional space enclosed by non-trivial operators at temporal index $k$.

\section{Numerical Results}
\begin{figure*}
\captionsetup[subfigure]{labelformat=empty}
    \subfloat[\label{subfig:comp_1D}]{}
    \subfloat[\label{subfig:del_rn_1D}]{}
    \subfloat[\label{subfig:comp_2D}]{}
    \subfloat[\label{subfig:del_rn_2D}]{}
     \includegraphics[width=\linewidth]{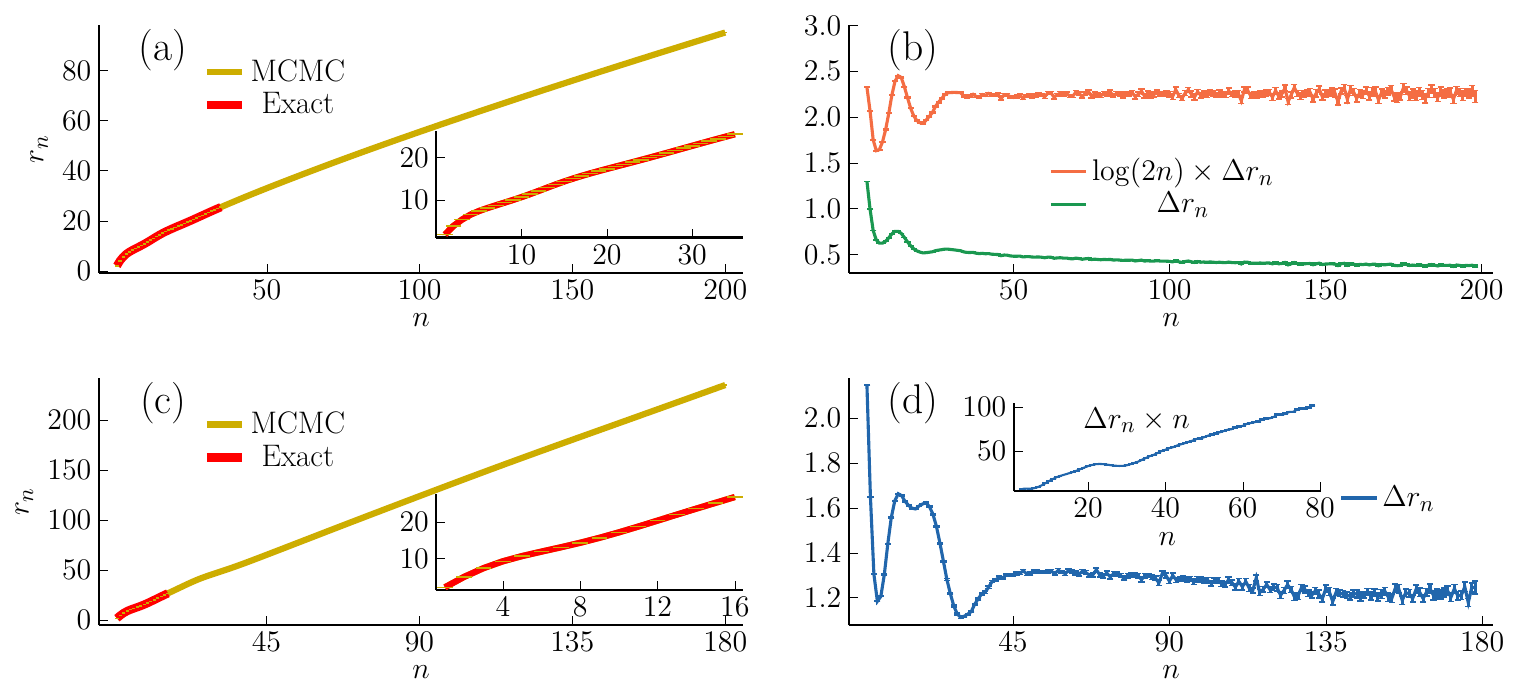}
     \vspace{0cm}
    \caption{\label{fig:moment_ratios} (a) $r_{n}$ vs $n$ for the $Z$ operator with $J=h^{Z}=h^{X}=1$ in 1D. The inset zooms in on the comparison of our MCMC calculations for the first 35 moments with brute-force exact numerics. (b) $\Delta r_{n}$ vs $n$ for 1D. This asymptotically falls to zero. Multiplying by $\log(2n)$ recovers a constant plateau.  (c) Same as (a) for 2D. The inset compares the first 16 moments against exact computation. (d) $\Delta r_{n}$ vs $n$ for 2D.}
\end{figure*}

In the following, we present Monte Carlo data for the $Z$ operator dynamics associated with the spectral function $\Phi_{ZZ}(\omega)=\int d\tau e^{i \omega \tau} \langle Z(\tau=it)Z(0)\rangle$ in the mixed field QIM with the microscopic parameters $h^X=h^Z=J=1$  in 1D and 2D and contrast them against brute force computations. The brute force calculations are performed by explicitly summing over all possible operator paths contributing to the moments, resulting in an unfavourable super-exponential scaling with the expansion order $n$. We begin our presentation by examining the 1D case. We recall that due to restricted dynamics in 1D,  the linear growth of $r_{n}$ receives a logarithmic correction such that $r_n\sim n/\log{n}$~\cite{araki69,Bouch,ADHH,Abanin_2015,Parker_2019}.  In \cref{subfig:comp_1D}, we depict $r_n$ vs $n$. We first note the excellent agreement with brute force calculation as shown in the inset, serving as a non-trivial benchmark of our approach. 
Furthermore, we observe that the large $n$ growth of $r_n$ deviates from a linear trend. To pin down the source of this deviation and check whether it can be attributed to the aforementioned logarithmic correction in \cref{subfig:del_rn_1D}, we plot $\Delta r_n \times \log(2n)$, where $\Delta r_n$ is a numerical derivative extracted using the five-point stencil method.  This quantity is expected to reach a plateau at large $n$ with the logarithmic correction. Indeed, we find that only in the presence of the log multiplication, the desired plateau is obtained. By contrast, $\Delta r_n$ asymptotically decays to zero due to the $1/\log(2n)$ factor.  We note that this is the first numerically exact confirmation for the predicted log correction associated with the restricted 1D dynamics, as previous brute-force numerics cannot resolve this subtle functional dependence convincingly due to the limited number of accessible moments \cite{Parker_2019, Gemmer_2022, Noh_2021}. 

We now turn to study 2D systems. In \cref{subfig:comp_2D}, we plot the evolution of $r_n$  with the expansion order $n$. As before, we note the precise agreement with brute force computation, shown in the inset. Impressively, the stochastic approach allows access to the first 180 moments, as compared to brute force methods that are limited to the first 16 moments.  We observe a clear linear trend of $r_n$, which is the first extensive test of the operator growth hypothesis in a concrete model beyond 1D systems.

In \cref{subfig:del_rn_2D}, we plot $\Delta r_n$ vs $n$. We observe that $\Delta r_n$ settles into an intermediate false plateau $\Delta r_{n} \simeq 1.6J$ in the range $12\lesssim n \lesssim20$ before transitioning to its true asymptotic plateau of $\Delta r_{n} \simeq 1.24J$ that is truly established beyond $n\approx40$. Since a constant value of $\Delta r_{n}$ corresponds to an exponentially decaying spectral function with decay rate $\omega_{0}=\Delta r_{n}/2$, the two-plateau structure allows us to infer that the spectral density features a crossover between two exponential tails. If we suppose
\begin{equation} \label{eq:twotails}
    \Phi_{ZZ}(\omega) \sim \begin{cases} 
    e^{- |\omega| / \omega_1} & |\omega| \lesssim \omega_c \\ 
      e^{- |\omega| / \omega_2} & |\omega| \gtrsim \omega_c
    \end{cases}
\end{equation}
where $\omega_1$ and $\omega_2$ are the decay rates and $\omega_c$ is the crossover scale. Using a saddle point approximation on \cref{eq:mu2n}, we find that 

\begin{equation}
    \Delta r_n \approx \begin{cases}
        2 \omega_1 & n \lesssim n_1 = \omega_c /2 \omega_1 \\ 
        2 \omega_2 & n \gtrsim n_2 = \omega_c /2 \omega_2
    \end{cases}
\end{equation}
In particular, we have $n_1 \omega_1 \approx n_2 \omega_2 \approx \omega_c/2$. To show that our numerical results are consistent with \cref{eq:twotails}, we plot $n \Delta r_n$ in the inset of \cref{subfig:del_rn_2D}, and find $n_{1} \approx 20$ and $n_{2} \approx 30$. This gives a remarkably large $\omega_c = 2 n_1 \omega_1 \approx 32J$. We note that the brute force calculations, limited to the first $16$ moments, only capture the intermediate spectral decay rate $\omega_{1}$, missing the true asymptotic decay rate $\omega_{2}$ entirely.

Owing to the thermal nature of the dynamics, we expect the large $n$ behavior of  $r_{n}$ to be universal for typical local operators with a given set of model parameters. In \cref{app:x_op}, we indeed see this anticipated behavior for boundary $X$ operator, with $\log(2n) \Delta r_{n} \simeq 2.2(1)J$ and $\Delta r_{n} \simeq 1.26(4)J$ in 1D and 2D respectively, consistent with the plots here. We also find that the structure of the intermediate plateau as seen in 2D is not universal. Nevertheless, the asymptotic plateau value is reached only at a similar large value $n_2 \approx 40$. Correspondingly, the high-frequency tail of the spectral density is truly established only at a frequency that was numerically inaccessible before the present work.

\section{Statistical mechanics perspective of operator growth}
\begin{figure*}
     \includegraphics[width=\linewidth]{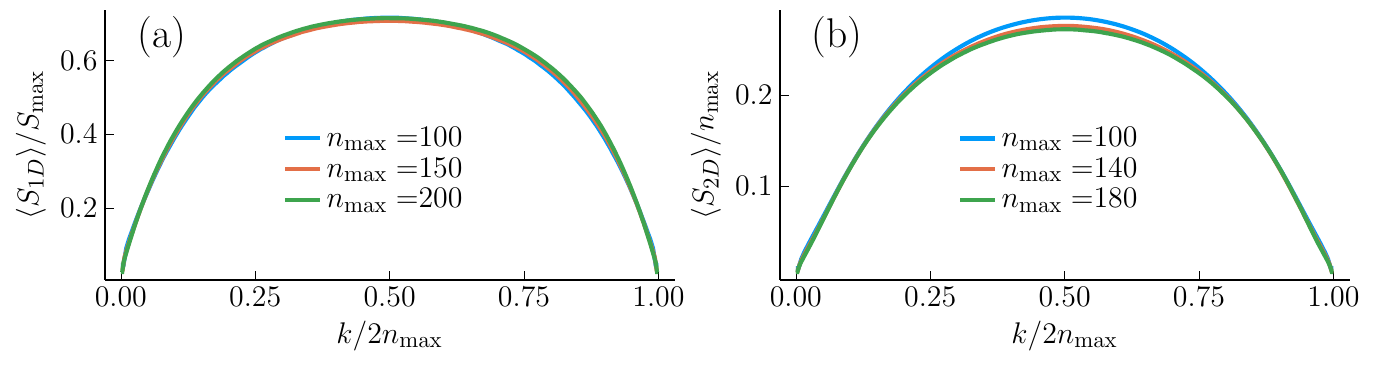}
\vspace{0.2cm}
\caption{\label{fig:collapse} Scaling collapse of the average operator support size (for the $Z$ operator), as function of $k$, in 1D (a), where $S_{\max} := n_{\max} / \log n_{\max}$, and in 2D (b).}
\end{figure*}
From the point of view of statistical mechanics, it is natural to examine geometric properties of the operator string configurations \cref{subfig:op_den}. In particular, configurations with large $n_{\max}$ describe the vicinity of the critical point $ \tau \nearrow \tau_*$ in the expansion \cref{eq:Ctau-exp} (in 1D $\tau_* = \infty$ because of the log-correction). Therefore, these large $n_{\max}$ configurations exhibit universal, scale-invariant behavior ubiquitous to continuous phase transitions.

To illustrate this point, we measured the average support size of operator strings against temporal index $k$ in configurations with an increasing range of maximal expansion order $n_{\max}$. The results, plotted in \cref{fig:collapse}, agree with the following scaling ansatz in the large $n_{\max}$ limit:
\begin{subequations}\label{eq:scalingforms}
\begin{align}
     &\left< S_{\text{2D}} \right> = n_{\max} f_{\text{2D}}(k / 2n_{\max}) , \label{eq:scaling2D} \\ 
    & \left< S_{\text{1D}} \right> =  \frac{n_{\max}}{\log (n_{\max})}  f_{\text{1D}}(k / 2n_{\max}) \label{eq:scaling1D}
\end{align} 
\end{subequations}
where $f_{\text{2D}}$ and $f_{\text{1D}}$ are scaling functions, which capture the evolution of the operator size growth to an $n_{\max}$-dependent maximum (and then decrease) in terms of rescaled time $ k / 2n_{\max}$ and size $ S / S_{\max}$. The scaling of $S_{\max} $ with respect to $n_{\max}$ is expected to be universal, and depend on the dimension. In 2D and higher dimensions, $S_{\max} \sim n_{\max}$. This is the only asymptotics compatible with the universal moment growth in ~\cref{eq:large_mu}. Indeed, the slow convergence to the scaling collapse echos the late establishment of the universal growth observed above.  In 1D, there is a log-correction $ S_{\max} = n_{\max} / \log(n_{\max})$. This has the same geometric origin as the log-correction to the moment growth: the operator can only grow at the two extremities, which is a severe entropy penalty. Therefore the dominant contributions are a compromise: the operator grows less in order to ``scramble'' more in the bulk, and $S_{\max}$ is the optimal scaling form of the maximal size~\cite{araki69, Bouch}. Finally, the scaling functions $f_{\text{2D}}$ and $f_{\text{1D}}$ are independent of the operator, which corresponds to the boundary condition of the partition sum. To what extent they depend on the Hamiltonian is an interesting question.

\section{Discussion}
We conclude our presentation by flagging several research directions motivated by our results. Our approach can be extended to models whose associated Liouvillian dynamics are expressed as sign problem-free sums. The quantum XY model is a notable example \cite{Cao_2021}, which crucially admits a global $\text{U}(1)$ symmetry and hence can provide access to transport properties of conserved charges \cite{Lindner_2010, Bhattacharyya_2023}. The presence of disorder is also compatible with our method, it will be interesting to investigate signatures of  Many-Body Localization in the evolution of operator strings. Another important direction is extracting real frequency dynamical spectral functions away from the high-frequency limit, as can, in principle, be via Lanczos coefficients expressed in terms of moment ratios \cite{Vishwanath_1990, Viswanath_2008, Khait_2016}. We note that the latter might suffer from high sensitivity to statistical noise and may require an improvement on current algorithms. Nevertheless, seeing how our Monte Carlo method fares against the existing techniques \cite{Rakovszky_2022,Wang_2023,Uskov_2024} will be an interesting line of enquiry. Lastly, we expect the support size scaling functions to be universal and valid for a generic class of chaotic Hamiltonians, and predicting them theoretically is an intriguing open question. We leave the aforementioned possibilities to future efforts.

\begin{acknowledgements}
We thank Ehud Altman, Assa Auerbach, Thomas Scaffidi and Daniel Parker for fruitful discussions and comments. S.G. acknowledges support from the Israel Science Foundation (ISF) Grant no.
586/22 and the US–Israel Binational Science Foundation (BSF) Grant no.  2020264. U.B. acknowledges support from the Israel Academy of Sciences and Humanities through the Excellence Fellowship for International Postdoctoral Researchers. Computational resources were provided by the Intel Labs Academic Compute Environment and the Fritz Haber Center for Molecular Dynamics, Hebrew University. X.C. acknowledges support from a France-Berkeley Fund grant (Project \#24\_2023).
\end{acknowledgements}

\appendix
\section{The Monte Carlo Algorithm}\label{app:MC}
In this section, we outline our Markov chain Monte Carlo (MCMC) sampling algorithm. We first discuss the re-weighting procedure used in \cref{eq:Z}. The weights $\{ w_{n} \}$ may be initialized as  $w_{n}=1/(J^{2n}(2n)!)$. This choice is useful as it is an upper bound to the radius of converge of the $\mu_{2n}$ sequence \cite{Cao_2021}. In practice one only needs to keep track of weight ratios between consecutive $n$ sectors $w_{n}/w_{n+1}$. The reweighting procedure works iteratively.  At each step, we take $N$ samples and determine the number of times $c_{n}$ that each sector is visited by the MCMC. We define $c_{n}$ as:
\begin{equation}
\label{sm_eq:wang_landau_counts}
c_{n}=\sum_{i=1}^{N}\delta_{n,n_{i}} \,.
\end{equation}
For a flat distribution, one expects $c_{n} \sim (n_{\max}+1)/N$. To bias our distribution in this direction, we update the weights by setting $w_{n} \rightarrow w_{n} N/(c_{n} (n_{\max}+1))$. The corresponding stored weight ratios get updated as: $w_{n}/w_{n+1} \rightarrow (w_{n}/w_{n+1})\times (c_{n+1}/c_{n})$. This assignment penalizes sectors with $c_{n} > (n_{\max}+1)/N$ by reducing their weight and reinforces weights for sectors with $c_{n} < (n_{\max}+1)/N$. This procedure is carried out repeatedly until an approximately flat distribution is obtained.

We now move our discussion towards the Monte Carlo moves. The moves are designed to ensure ergodicity while only making local changes to the configuration space. Below, we discuss the three classes of moves: \texttt{add-drop, swap} and \texttt{interact}. For the rest of this section, we will work with a $d + 1$ dimensional lattice with length $L$ and $2n_{\max} + 1$ temporal slices. The temporal axis is pictured in the vertical direction. While the schematic diagrams show the moves on a $d=1$ lattice, they generally also apply to higher dimensional cases. We will point out subtleties pertaining to $d>1$ dimensions when appropriate.
\subsection{Add-drop}
\label{sm_subsec:add_drop}
\begin{figure}[h]
\captionsetup[subfigure]{labelformat=empty}
    \subfloat[\label{sm_subfig:add_drop_z}]{}
    \subfloat[\label{sm_subfig:add_drop_x}]{}
    \subfloat[\label{sm_subfig:add_drop_zz}]{}
     \includegraphics[width=\columnwidth]{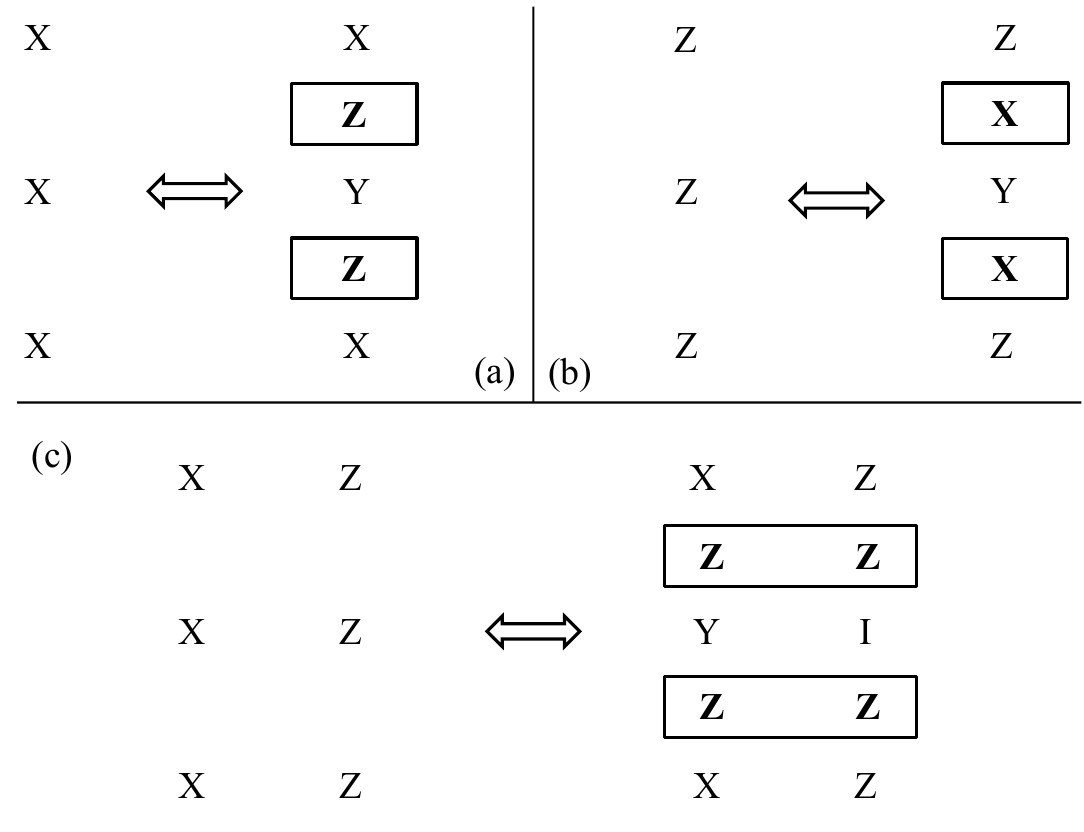}
     \vspace{0cm}
    \centering \caption{\label{sm_fig: add_drop} Illustration of the \texttt{add-drop} moves with (a) $\mathcal{L}_{Z}$ Liouvillians (b) $\mathcal{L}_{X}$ Liouvillians (c) $\mathcal{L}_{ZZ}$ Liouvillians. The temporal direction is vertical.} 
\end{figure}
The \texttt{add-drop} move is designed to allow transitioning between the different $n$ sectors by adding and removing a pair of identical local Liouvillians. As illustrated in Fig.~\ref{sm_fig: add_drop}, the move is defined as follows:
\begin{enumerate}
\item Randomly choose a time slice with uniform probability.
\item If the local Liouvillians occupying both sides of the time slice are identical, propose to remove both.
\item Else, if the Liouvillian slots on both sides are unoccupied, randomly choose a site with uniform probability and propose to add a pair of identical Liouvillians onto the unoccupied slots at that site.
\end{enumerate}

We now determine the transition probabilities. Let $i$ be a configuration in the $n-1$ sector with the $(k-1)$-th and $k$-th Liouvillian slots unoccupied. Let $f$ be a configuration in the $n$ sector that is identical to $i$ except with these slots occupied at site $r$ by Liouvillians $\mathcal{L}_{r_{k-1},a_{k-1}} = \mathcal{L}_{r_k,a_k} =: \mathcal{L}_{r,a}$. The two configurations are thus related by an add-drop move. Based on \cref{eq:Z}, detailed balance imposes the constraint:

\begin{align}
\label{sm_eq:add_drop_detailed_balance}
& \frac{T_{i\rightarrow f}A_{i \rightarrow f}}{T_{f\rightarrow i}A_{f \rightarrow i}}=\frac{P_{f}}{P_{i}} \nonumber \\
& =\frac{w_{n}(O_{k-1}|\mathcal{L}_{r,a}|O_{k})(O_{k}|\mathcal{L}_{r,a}|O_{k+1})(2n)(2n-1)}{w_{n-1}(2n_{\max}-2n+2)(2n_{\max}-2n+1)} \,.
\end{align}
Here, $P_{i}, \space P_{f}$ are associated probabilities for states $i,f$; $T_{i\rightarrow f}, \space T_{f \rightarrow i}$ are probabilities to propose the add/drop transitions and $A_{i\rightarrow f}, \space A_{f \rightarrow i}$ are the acceptance probabilities for the moves. Now, proposing an \texttt{add} operation involves choosing a time slice, followed by choosing a site and then choosing a Liouvillian type; proposing a \texttt{drop} amounts to simply choosing a time slice. Therefore the proposition probabilities are 
\begin{align}
    &T_{i\rightarrow f}=\frac{1}{2n_{\max}+1} \times \frac{1}{L^{d}} \times \frac{(O_{k-1}|\mathcal{L}_{r,a}|O_{k})(O_{k}|\mathcal{L}_{r,a}|O_{k+1})}{\sigma} \,,\, \nonumber \\
&T_{f\rightarrow i}=\frac{1}{2n_{\max}+1} \,.
\end{align}
where we denoted $\sigma=4(d J^{2}+(h^{X})^{2}+(h^{Z})^{2})$ as the sum of weights of all possible Liouvillians at a site. Therefore, the following ``Metropolis'' acceptance rates satisfy detailed balance:
\begin{align}
&A_{i\rightarrow f}=\text{min}\left(1,\frac{w_{n}L^{d}\sigma(2n)(2n-1)}
{w_{n-1}(2n_{\max}-2n+2)(2n_{\max}-2n+1)}\right)\,,\, \nonumber \\ 
&A_{f\rightarrow i}=\text{min}\left(1,\frac{w_{n-1}(2n_{\max}-2n+2)(2n_{\max}-2n+1)}{w_{n}L^{d} \sigma(2n)(2n-1)}\right) \,.
\end{align}
\subsection{Swap}
\begin{figure}[t]
     \includegraphics[width=\columnwidth]{
     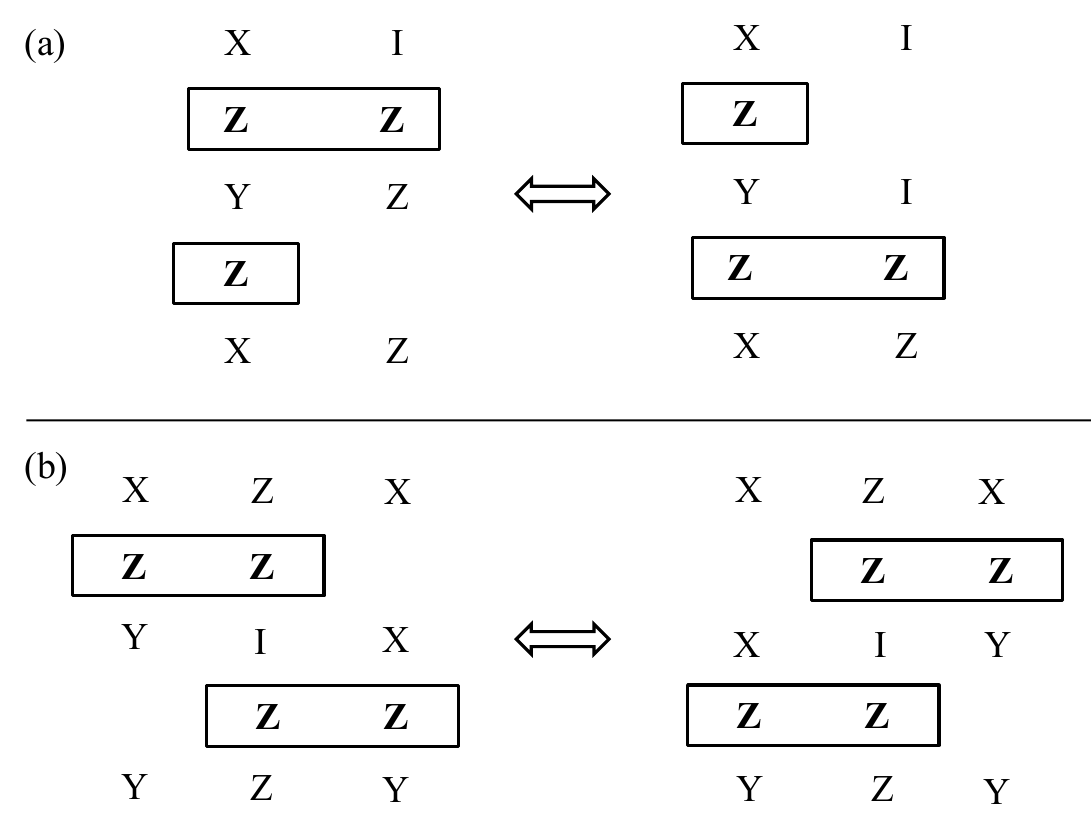}
     \vspace{.5cm}
    \centering \caption{\label{sm_fig:scramble} Illustration of the \texttt{swap} move. We show two examples of Liouvillians that commute with each other: (a) $\mathcal{L}_{Z}$ and $\mathcal{L}_{ZZ}$ Liouvillians sharing a site. (b) Two $\mathcal{L}_{ZZ}$ Liouvillians sharing just one site. In addition to (a) and (b), identical Liouvillians (those of the same kind, on the same site; and along the same axis in case of $\mathcal{L}_{ZZ}$) and those on different sites with no overlap, trivially commute.} 
\end{figure}
The \texttt{swap} move permutes through different configurations in the same $n$ sector using the commutation property of Liouvillian pairs, see \cref{sm_fig:scramble} for an illustration. Before discussing the steps, we first define the notion of ``ceiling" and ``floor". The ceiling (floor) of a Liouvillian is the temporally closest Liouvillian above (below, respectively) it which doesn't commute with it. If a non-commuting Liouvillian is lacking, the ceiling/floor is then defined to be the initial/final temporal boundary. Now, the \texttt{swap} steps are as follows:
\begin{enumerate}
\item Randomly, with uniform probability, choose a Liouvillian.
\item Randomly, with uniform probability, choose a second Liouvillian slot amongst those lying between the ceiling and floor of the first Liouvillian.
\item If the newly chosen slot is empty, remove the Liouvillian from its existing slot and place it in the new slot.
\item Otherwise, that is, if the slot is occupied by another Liouvillian, check whether the ceiling and floor of this second Liouvillian lie at an intermediate temporal index between the indices of the two Liouvillians. If that is not the case, swap the two Liouvillians.
\end{enumerate}
Since the \texttt{swap} moves involve no weight change and everything is sampled uniformly, it satisfies detailed balance automatically, that is, with $100 \%$ acceptance rate.

\subsection{Interact}
\label{sm_subsec:interact}

\begin{figure}[t]
     \includegraphics[width=\columnwidth]{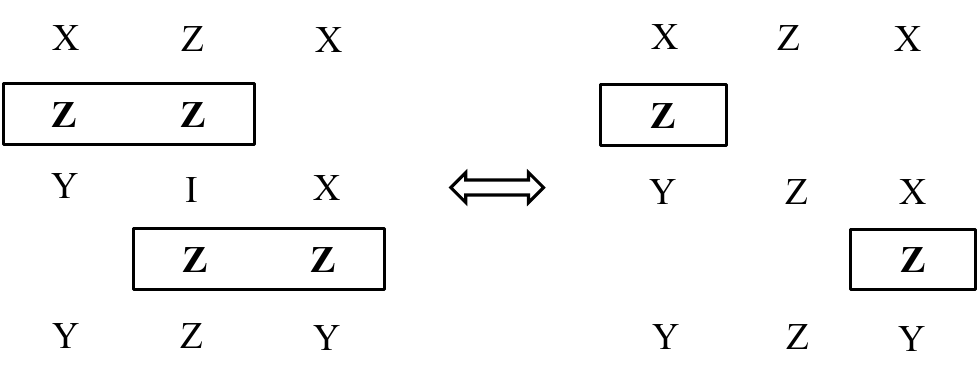}
     \vspace{.5cm}
    \centering \caption{\label{sm_fig:interact} Illustration of the \texttt{interact} move. Two $\mathcal{L}_{ZZ}$ Liouvillians sharing only one site may ``interact" to give a pair of $\mathcal{L}_{Z}$ Liouvillians situated on next to nearest neighbour sites.} 
\end{figure}
The \texttt{add-drop} move only allows Liouvillians to exist on a site in pairs. This can however miss configurations with an odd number of bond Liouvillians on a site. To enable this we introduce the \texttt{interact} move, as illustrated in \cref{sm_fig:interact}. It is to be noted that for $d>1$, the two interacting bonds can be along different spatial axes. In such cases, two different $\mathcal{L}_{ZZ}$ pair configurations can map to the same $\mathcal{L}_{Z}$ pair configuration. This needs to be accounted for in detailed balance. 

The interaction is implemented as:
\begin{enumerate}
\item Randomly choose a time slice with uniform probability.
\item If the Liouvillian on both sides of the time slice are of type $\mathcal{L}_{ZZ}$ and share exactly one site, check if the shared site has a $Z$ or $I$ operator on it. If so, replace the $\mathcal{L}_{ZZ}$ Liouvillians with $\mathcal{L}_{Z}$ Liouvillians as in \cref{sm_fig:interact}.
\item If the Liouvillian slots on both sides host $\mathcal{L}_{Z}$ Liouvillians placed at next to nearest neighbour sites, check if their common nearest neighbouring site has a $Z$ or $I$ operator. If this is the case, replace the $\mathcal{L}_{Z}$ Liouvillians with $\mathcal{L}_{ZZ}$ Liouvillians as in \cref{sm_fig:interact}.
\end{enumerate}
Now, we derive the transition probabilities. Let $i$ be a configuration with $\mathcal{L}_{ZZ}$ Liouvillians on the $k-1$-th and $k$-th time slots sharing a single site $r$. Let $f$ be the configuration related to $i$ by the \texttt{interact} move, hosting two $\mathcal{L}_{Z}$ Liouvillians. Now, detailed balance implies:
\begin{equation}
\label{sm_eq:interact_detailed_balance}
\frac{T_{i\rightarrow f}A_{i \rightarrow f}}{T_{f\rightarrow i}A_{f \rightarrow i}}=\frac{P_{f}}{P_{i}}=\frac{(h^{Z})^2}{J^{2}}
\end{equation}
where $P_{i}, \space P_{f}$ are the respective weights, $T_{i\rightarrow f}, \space T_{f \rightarrow i}$ are proposition probabilities and $A_{i\rightarrow f},\space A_{f \rightarrow i}$ the acceptance probabilities. We claim that the proposition probabilities are as follows:
\begin{equation}
\label{sm_eq:interact_tif}
T_{i\rightarrow f}=\frac{1}{2n_{\max}+1} \,,\, T_{f\rightarrow i}= \frac{1}{(2n_{\max}+1) \alpha} \,,
\end{equation}
where $\alpha=2$ if the $\mathcal{L}_{ZZ}$ Liouvillians in configuration $i$ are along different spatial directions (otherwise, $\alpha=1$).
This is because, proposing a $\mathcal{L}_{ZZ}\to \mathcal{L}_{Z}$ \texttt{interact} move amounts to choosing a time slice. Meanwhile, to specify a $ \mathcal{L}_{Z} \to \mathcal{L}_{ZZ}$ \texttt{interact} move, we need to choose a time slice, and then pick a common neighboring site among the $\alpha$ choices. Therefore, the following acceptance rates fulfill detailed balance:
\begin{equation}
\label{sm_eq:interact_aif}
A_{i\rightarrow f}=\text{min}\left(1,\frac{(h^{Z})^2}{\alpha J^{2}} \right) \,,\,
A_{f\rightarrow i}=\text{min}\left(1,\frac{\alpha J^{2}}{(h^{Z})^{2}} \right) \,.
\end{equation}

\section{ $X$ Correlation Function}
\label{app:x_op}
\begin{figure}[t]
\captionsetup[subfigure]{labelformat=empty}
    \subfloat[\label{sm_subfig:comp_1D}]{}
    \subfloat[\label{sm_subfig:del_rn_1D}]{}
    \subfloat[\label{sm_subfig:comp_2D}]{}
    \subfloat[\label{sm_subfig:del_rn_2D}]{}
     \includegraphics[width=\columnwidth]{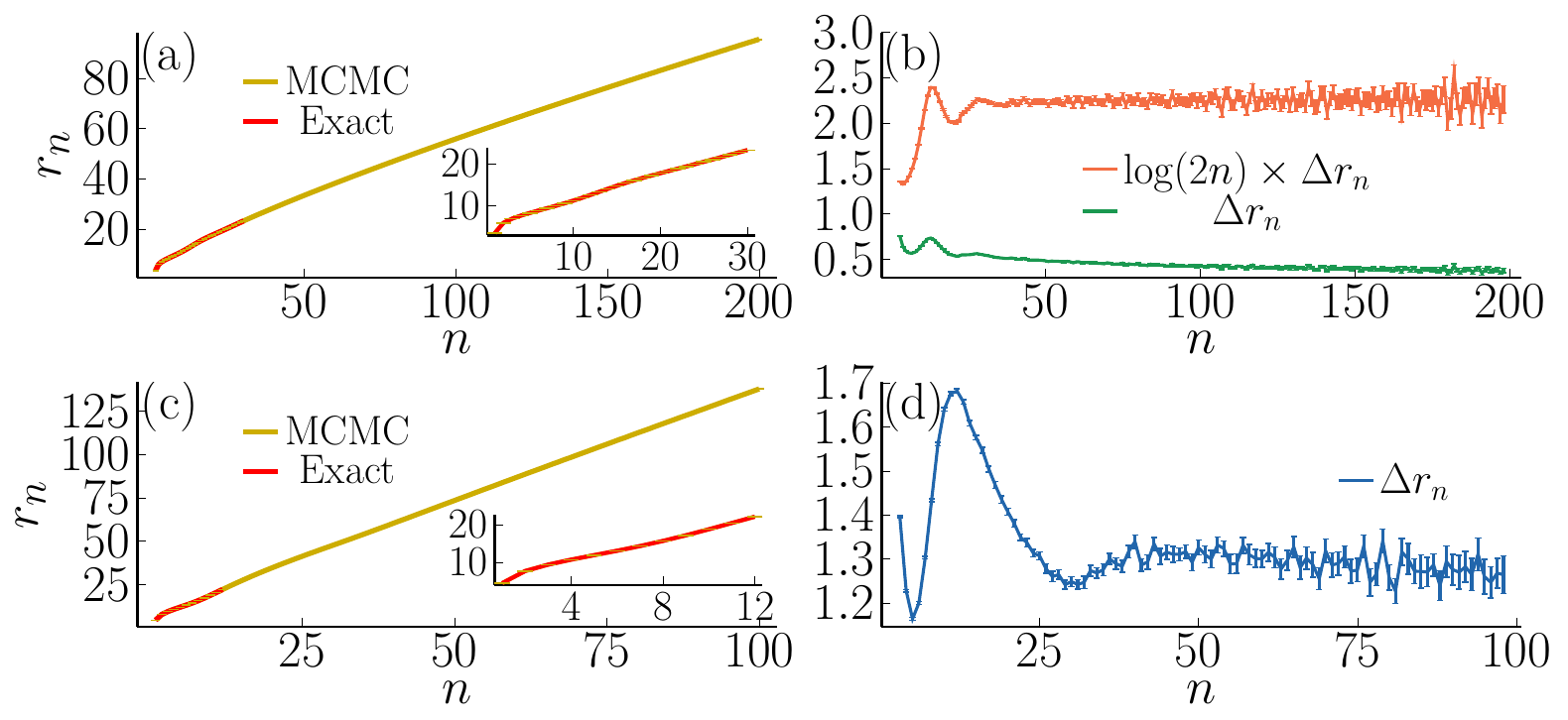}
     \vspace{0cm}
    \caption{\label{sm_fig:sigmax_moment_ratios} (a) $r_{n}$ vs $n$ for the $X$ operator with $J=h^{Z}=h^{X}=1$ in 1D. The inset focuses in on the first 30 moments, comparing our MCMC with exact numerics. (b) $\Delta r_{n}$ and $\log(n)\Delta r_{n}$ vs $n$ for 1D. (c) Same as (a) for 2D. The inset has the first 12 moments. (d) $\Delta r_{n}$ vs $n$ for 2D. }
\end{figure}

The statistical nature of infinite temperature operator growth dynamics in non-integrable systems like the mixed field QIM implies that local operators should typically give rise to the same generic behaviour at a large enough order $n$. Based on the Operator Growth Hypothesis \cite{Parker_2019}, one expects local operators to asymptotically grow linearly with the same slope. To check this, we compute the same plots as  
Fig.~3 in the main text but for the $X$ operator in \cref{sm_fig:sigmax_moment_ratios}.  We see that for 1D $\log(2n)\Delta r_{n}$ vs $n$ again gives us a constant curve $\log(2n)\Delta r_{n} \simeq 2.2(1)J$, the same as the $Z$ operator in the main text. Similarly, in 2D we again get a plot consistent with linear growth with slope $\simeq1.26(4)J$. 
Another point to note is the non-universal nature of the intermediate plateau in the 2D $\Delta r_{n}$ vs $n$ plot. We see that here the plateau is much narrower than the one for the $Z$ operator as shown in the main text.
\begin{figure}[t]
     \includegraphics[width=0.9\columnwidth]{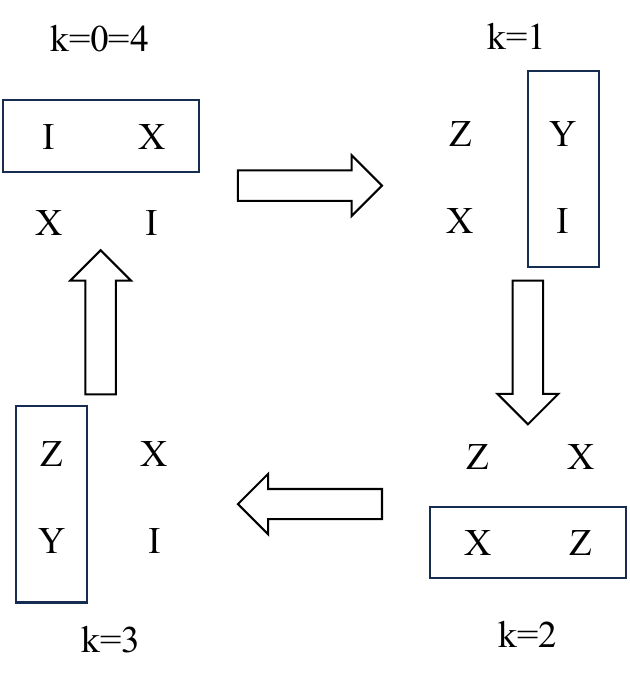}
     \vspace{.5cm}
    \centering \caption{\label{sm_fig:plaquette}An example of a configuration in the $n=2$ sector that requires the \texttt{interact} move to be generated. Temporal slices of the 2D lattice are indexed by k. The first temporal slice is identical to the last because of the fixed boundary conditions.  The $\mathcal{L}_{ZZ}$ Liouvillians, shown as rectangular boxes, transform local Pauli operators between slices.}
\end{figure}
\section{Additional move for the TFIM Limit}

\begin{figure}[t]
     \includegraphics[width=\columnwidth]{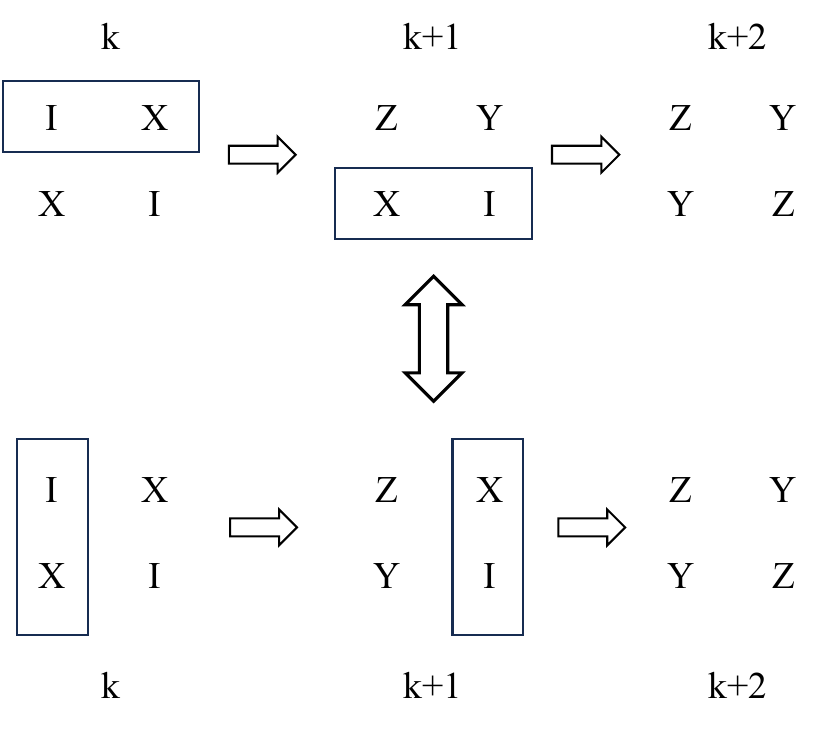}
     \vspace{.5cm}
    \centering \caption{\label{sm_fig:rotate} Section of two configurations connected by the \texttt{rotate} move on a 2D lattice. k corresponds to the temporal index.  Note that both configurations have the same temporal boundaries, implying the local nature of the move.}
\end{figure}

In the $h^{Z}=0$ limit, the \texttt{interact} move breaks down. This is not a problem in 1D as configurations with an odd number of identical $\mathcal{L}_{ZZ}$ Liouvillians also require the presence of $\mathcal{L}_{Z}$ Liouvillians and thus do not exist in this limit. For $d>1$ it is possible for configurations to have an odd number of identical  $\mathcal{L}_{ZZ}$ Liouvillians without $\mathcal{L}_{Z}$ Liouvillians. \cref{sm_fig:plaquette} shows a simple example with $\mathcal{L}_{ZZ}$ Liouvillians \emph{looping} around a plaquette, satisfying the boundary conditions. Note that each site hosts just one $\mathcal{L}_{ZZ}$ Liouvillian of a kind. One can generate larger \emph{loops} by removing $\mathcal{L}_{ZZ}$ pairs on the common edges of overlapping smaller \emph{loops} using \texttt{add-drop}.

\begin{figure}[bt]
     \includegraphics[width=\columnwidth]{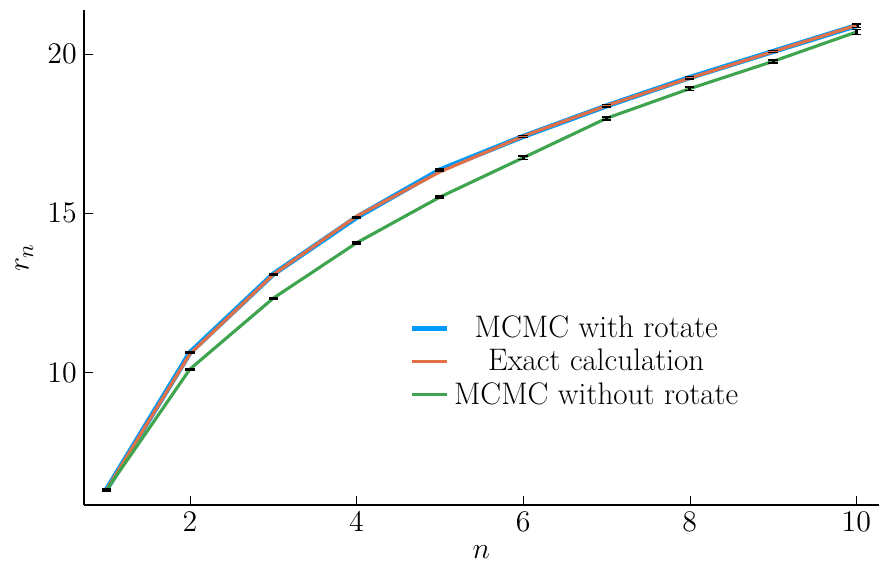}
     \vspace{.5cm}
    \centering \caption{\label{sm_fig:rotate_benchmark} Comparing our MCMC equipped with and without the \texttt{rotate} move against exact computations for the 2D TFIM with the same boundary operator as in \cref{sm_fig:plaquette}.}
\end{figure}

To access this part of the configuration space in absence of \texttt{interact} we propose the \texttt{rotate} move. \cref{sm_fig:rotate} shows that it is possible to rotate two parallel $\mathcal{L}_{ZZ}$ Liouvillians sharing a plaquette to get another pair of parallel $\mathcal{L}_{ZZ}$ Liouvillians on the same plaquette while maintaining the same boundaries. Detailed balance for this move should be trivial as the involved configurations have the same weight and thus the Metropolis acceptance rate is $100\%$. It can be seen that a combination of \texttt{add-drop} and \texttt{rotate} can generate the configuration in \cref{sm_fig:plaquette}. We benchmark this move against exact numerics in \cref{sm_fig:rotate_benchmark}.

\bibliography{References} 
\end{document}